\documentclass[conference]{IEEEtran}
\usepackage{balance}

\usepackage[colorlinks, citecolor=blue]{hyperref}

\usepackage{cite}

\ifCLASSINFOpdf
\else
\fi
\usepackage{amsmath}
\usepackage{mathtools,amssymb}
\usepackage{algorithmic}

\usepackage{array}

\newcommand{\eg}{{\it e.g., }}
\newcommand{\etal}{{\it et~al., }}
\newcommand{\ie}{{\it i.e., }}
\begin{document}
%
\title{Serverless Edge Computing for Green Oil and Gas Industry}


\author{
	{\bfseries Razin Farhan Hussain$^1$, Mohsen Amini Salehi$^1$, and Omid Semiari$^2$}\\
	$^1${razinfarhan.hussain1, amini@louisiana.edu}\\
	$^1$High Performance Cloud Computing (HPCC) Laboratory, School of Computing and Informatics,\\ University of Louisiana at Lafayette, Lafayette, LA, USA\\
	$^2${osemiari@georgiasouthern.edu}\\
	$^2$Department of Electrical and Computer Engineering, Georgia Southern University, Statesboro, GA, USA\\
}


%


\maketitle

\begin{abstract}
Escalating demand of petroleum led the Oil and Gas (O\&G) industry to extend oil extraction operation in the remote reservoirs. Oil extraction is a fault intolerant process where the maximum penalty is disaster impacting the environment seriously. Therefore, efficient and nature-friendly green oil extraction is a challenging operation, especially with location constrained in accessing the sites. To overcome these challenges and protect the environment from pollution, smart oil fields with numerous sensors (\eg for pipeline pressure, gas leakage, air pollution) are established to achieve clean O\&G extraction. Conventionally, cloud datacenters are utilized to process the generated data. High-latency satellite communication are used for data transfer, which is not suitable for time-sensitive operations/tasks. To process such latency-sensitive tasks, edge computing can be a suitable candidate, however, their computational power goes downhill at disaster time due to surge demand of many coordinated activities. Therefore, we propose green smart oil fields that operate based on edge computing. To overcome shortage of resources and rapid deployment of the edge computing systems, we propose to use lightweight serverless computing on a federation of edge computing resources from nearby oil rigs. Our solution coordinates urgent coordinated operations/tasks to prevent disasters in oil fields and enable the idea of green smart oil fields. Evaluation results demonstrate the efficacy of our proposed solution in compare to conventional solutions for smart oil fields.
\end{abstract}



%

\IEEEpeerreviewmaketitle
\section{Introduction}
Petroleum has been unquestionably one of the most important drivers of the world economy in recent decades. Due to high demand of petroleum, Oil \& Gas (O\&G) industry is expanding the extraction operations in remote and adverse locations (\eg Golf of Mexico, Persian Gulf, West Africa) \cite{mathieson2007forces} where giant reservoir exist, hence, several oil extraction sites are constructed within a short distance. The complex oil extraction process requires high reliability and extra safety measures to protect the surrounding environment. Moreover, governments (\eg U.S. environmental protection agency (EPA)) also enforcing regulations on O\&G industries to reduce the adverse impact of oil extraction on the environment.


To protect the environment from disasters~\cite{White20303} that can occur due to flaws in oil extraction process, oil fields are geared up with many cyber devices (\eg sensors and actuators) and the concept of \emph{smart oil field} has emerged with green oil extraction motto. Smart oil fields utilize various sensors (temperature, Hydrogen Sulphide (H2S) gas emission, pipeline pressure, air pollution) which ​gather a large volume of data (up to two Terabytes per day~\cite{sof2}). To enable ultra-reliable and flawless green oil extraction these sensor generated datasets need to be analyzed and used in a real-time manner. The need for smart oil fields has been emphasized by both industry \cite{sof3,sof2,sof5} and academia \cite{sof1,LIU20112611,sof12W} to improve the efficiency of oil field and save the environment from pollution. However, existing smart oil field solutions cannot meet the requirements of \emph{remote} oil fields for two specific reasons: (1) Lack of reliable and fast communications infrastructure to access to onshore management teams; (2) High cost of operations by human resources to perform real-time inspection and monitoring. 

Current remote smart oil field solutions utilize satellite communication to cloud datacenters which is known to be unstable and imposes a significant latency. Hence, the goal of this study is to enable the idea of smart oil fields in remote sites for nature-friendly oil extraction. This research mainly focus on exploiting edge computing system in serverless manner for remote oil fields with unstable and weak connectivity to datacenter to enable the idea of Green O\&G Industry. To handle latency-sensitive tasks, specifically, during a disaster when there is a surge for real-time computation, the edge computing system plays a vital role due to its locality to end user. However, the main challenges of utilizing edge computing are the resource constrained nature of edge nodes and difficulty of configuration and maintenance in remote areas. To overcome these problems and enable a robust system against the surge in demand, we harness the edge devices located in nearby oil rigs and propose a technique to federate them in an on-demand manner. For ease of federation and configuration, we propose to use serverless computing paradigm on the edge computing systems. 

Federation of serverless edge computing systems can alleviate the shortage of resources, however, it introduces new challenges of processing tasks in the federated environment. As such, the research problem is \emph{how to allocate surge requests to a serverless federated edge computing system, considering uncertainties exist during disaster times in these environments?}
 
To address this problem, we introduce a \emph{service balancer} for each edge nodes that provides Quality of Service (QoS) by considering the federation of edge nodes. Accordingly, the service balancer decides whether to allocate the arriving service request (aka task) locally (\ie on the receiving edge) or on a neighboring edge. Then, we propose a probabilistic model and develop a resource allocation heuristic for the service balancer to utilize the edge federation. As the serverless edge computing system has a central role in the smart oil field, its ability to cope with surge loads results in having a green energy production and O\&G industry. 

In summary, The contributions of this paper are as follows:
\begin{itemize}
	\item Proposing a federation of serverless edge computing system to enable green oil extraction utilizing a robust resource allocation scheme with minimum connectivity to onshore cloud datacenter.
	\item Developing a model to capture the uncertainties exist in the federated edge environment.
	\item Analyzing the performance of the federated edge computing system under various oversubscribed conditions.​
\end{itemize}

The rest of the paper is organized in the following manner. Section 2 presents the system model. Section 3 and 4 represent system architecture and task distribution in edge federation respectively. Section 5 demonstrates the performance evaluation and experiments. Section 6 presents related work. Finally, Section 7 concludes the paper with some future directions for exploration.

\section{System Model }
In our system model, we consider utilizing edge machines which include storage, computational power, and communication capacity. The edge machines can be placed on the platform of oil rig above the water surface or can be mount on a floating boat near the site. Due to hardware limitation, edge machines are more appropriate for the real-time urgent task processing which typically has shorter deadline and delay-sensitive in nature. We consider utilizing serverless edge platform to facilitate management of resource allocation and optimal placement of smart oil field micro services (\ie database service, image processing). The sensors (\eg temperature, flow rate, tank level, gas leakage sensors) in a smart oil field generate a large amount of diverse data which is utilized by different applications. Accordingly, we classified these applications as task type (service type). These task types can vary from processing surveilled images (taken by unmanned aerial vehicles (UAVs) or embedded cameras) for detecting oil spill anomalies~\cite{liu2011oilUAV}; Analyzing large volume of data, streamed by sensors, to predict the oil spill spread direction and quantity\cite{fingas2011oil}. Because the task types have various computational demands, they need processing machines with different characteristics (\ie heterogeneous machines). This form of HC systems is known as inconsistently heterogeneous systems~\cite{MAHESWARAN1999107}. We assume different machine provides various micro services in serverless manner. Upon arrival of a task of type $i$ to a compute node $j$, it is assigned an individual deadline based on its arrival time and the end-to-end delay it can tolerate. The deadline can be defined as:
$\delta_{ij}$ = $arr_{i}$ + $\beta\times avg_i$ + $\alpha\times d_{comm}$ + $\epsilon$, where $arr_{i}$ is arrival time of the task, $avg_i$ is average of completion time in all edge nodes, $d_{comm}$ is the communication latency, $\epsilon$ is system slack, $\beta$ is computing constant, and $\alpha$ is the communication constant.
During an incident, different task types have bust arrivals to the edge system and make the edge system oversubscribed. Hence, the system receives the number of tasks beyond its capacity. Consequently, some tasks are considered to miss their deadlines according to the level of oversubscription. For performance improvement of services in edge nodes, we assume utilizing serverless technology. 

\begin{figure}[h!]
	\centering	
	\includegraphics[scale=0.40]{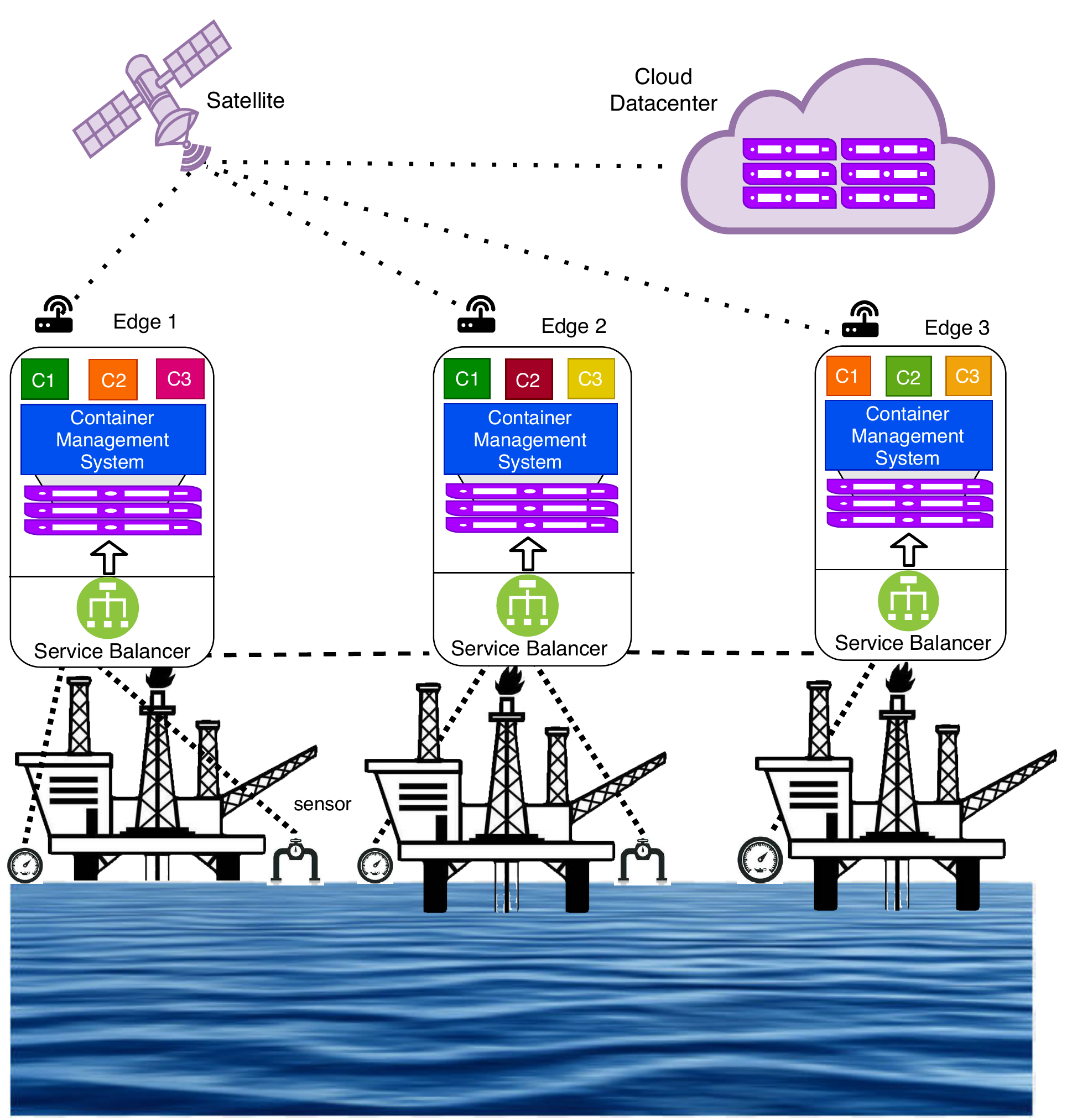}
	\caption{System Architecture of serverless edge computing in smart oil field where services are provided through containerized technology to sensor generated data.}
	\label{fig:systemArchitecture}
\end{figure}
\vspace{-5mm}
\section{System Architecture}
As stated in the system model, the proposed system architecture considers an edge device with a service balancer module in every oil extraction site as demonstrated in \ref{fig:systemArchitecture}. The system architecture includes two-tier of computing nodes where edge nodes are located in local or in the first tier and cloud data centers are in the second tier. Physical sensors (\ie flow rate sensor, pressure sensor, tank level sensor, gas sensor) of smart oil fields takes physical quantity and convert it to the electrical signal. The physical sensors include micro-controller for getting the readings which are defined as sensor units. The sensor unit has a communication interface (\eg Ethernet, Bluetooth, Wi-Fi, Zigbee) to communicate with the edge device. The sensor units send tasks to the first tier (edge device) of the architecture where incoming tasks are sorted in terms of latency (\ie latency intolerant, latency tolerant) by the service balancer. This architecture support the serverless edge platform stated in system model combining the benefits of edge with the computational and storage capabilities of cloud. 



\section{Task Allocation in Edge Federation}\label{solution}\vspace{-1mm}
Considering task allocation in edge federation, every edge node has a service balancer module which works in an immediate mode to allocate the incoming task to the appropriate computing node. We observe that the task completion time data from the historical record follows normal distribution according to the central limit theorem and can be applied to perform statistical modeling. Therefore, we consider every service balancer has access to this historical data which is stored in a matrix data structure and defined as Estimated Task Completion (ETC) time matrix. Every cell of this matrix represents normal distribution ($X \sim \mathcal{N}(\mu,\,\sigma^{2})$) of a particular task type. Considering the task processing in neighbor edge device of the federation, communication overhead has a significant impact in service time. Therefore, the transfer time of a task from a one service balancer to other edge nodes is captured in Estimated Task Transfer (ETT) time matrix. ETT matrix basically captures the communication uncertainty using normal distribution ($Y \sim \mathcal{N}(\mu,\,\sigma^{2})$). Both of the matrices are updated periodically to reflect the current situation which is utilized by the service balancer to estimate the probability of success.

\subsubsection{\textbf{Probabilistic Model}}
Upon arrival of a task $t$ of type $i$,  the service balancer calculates the probability ($P_i(t)$) of meeting the task's deadline $\delta_i$ in it's receiving edge node as well as it's neighbor edges. If receiving edge node is $j$, then the probability of success in  $j$ can be defined as:
\begin{equation}\label{eq:10}
P_i^j(E_c^j(t_i)<\delta_i) = P(Z<z), where~ z = (\delta_i - \mu_i^j)/\sigma_i^j
\end{equation}
where $E_c^j(t_i)$ is the estimated task completion time of task type $i$ in edge node $j$. $\mu_i^j$  and $\sigma_i^j$ are respectively the average and standard deviation of the considering distribution. In Equation~\ref{eq:10}, z score is used to standardize the normal distribution. For calculating the probability of neighbor edge nodes, the ETC matrix's normal distribution of receiving task type is convolved with its ETT matrix distribution which incorporates the communication overhead with computing overhead. After calculating the probability with resulting distribution for all the edge nodes with respect to task type, the task is allocated to the edge node that offers the highest probability.
\subsection{\textbf{Heuristic based on Probabilistic Model}}
\subsubsection{\textbf{Highest Probability of Success (HPS)}}
The heuristic allocates the arriving task based on its probability of success across the edge nodes. The service balancer utilizing $HPS$ heuristic estimates the probabilities which represent success of the arriving task to meet its deadline across the edge nodes. The $HPS$ heuristic chooses the maximum probability edge node to allocate receiving task.


\section{Evaluation}\label{experiment}
\subsection{\textbf{Experimental Setup \& Workload}}
For performance evaluation of our proposed model, EdgeCloudSim\cite{sonmez2017edgecloudsim} is considered which is a discrete event simulator specific to edge computing scenarios. Datacenters of EdgeCloudSim are considered as edge devices with limited computational capacity (1500-2500 MIPS) including 8 homogeneous cores. On the other hand, different data centers have different computational power (MIPS) which represents the heterogeneity across the edge nodes. A large cloud datacenter with massive computational power (40000 MIPS) is considered for non-urgent delay tolerant tasks. The WLAN bandwidth is set to 200Mbps and propagation delay is considered as 0.57 seconds\cite{goyal1998analysis} which occurs from satellite communication. 

EdgeCloudSim's default workload includes four different task types, among which two of them are urgent (\ie latency intolerant) and the other two are non-urgent (\ie latency tolerant). The execution time of a task is represented in Million Instructions per seconds(MIPs) which is sampled out as a normal distribution from an average value for a particular task type.
\subsection{\textbf{Baseline Heuristics}}
In this paper, we consider using two baseline heuristics. They are Minimum Expected Completion Time (MECT) and Success with Computational Certainty (SCC). MECT heuristic considers minimizing expected completion time for receiving task in edge nodes of federation whereas SCC tends to maximize the difference between the deadline and average completion of arriving tasks for allocation decision.
\subsection{\textbf{Results and Analysis}}
As deadline miss rate is the fundamental rubric for maintaining QoS, we evaluate our system based on this standard. To investigate the performance of our scheme with increased oversubscription level we increase the number of applications submitted to the system. The number of submitted applications increased from 50 to 250 which generate approximately 500 to 10000 tasks for the system. 
\begin{figure}[h!]
	\centering	
	\includegraphics[scale=0.54]{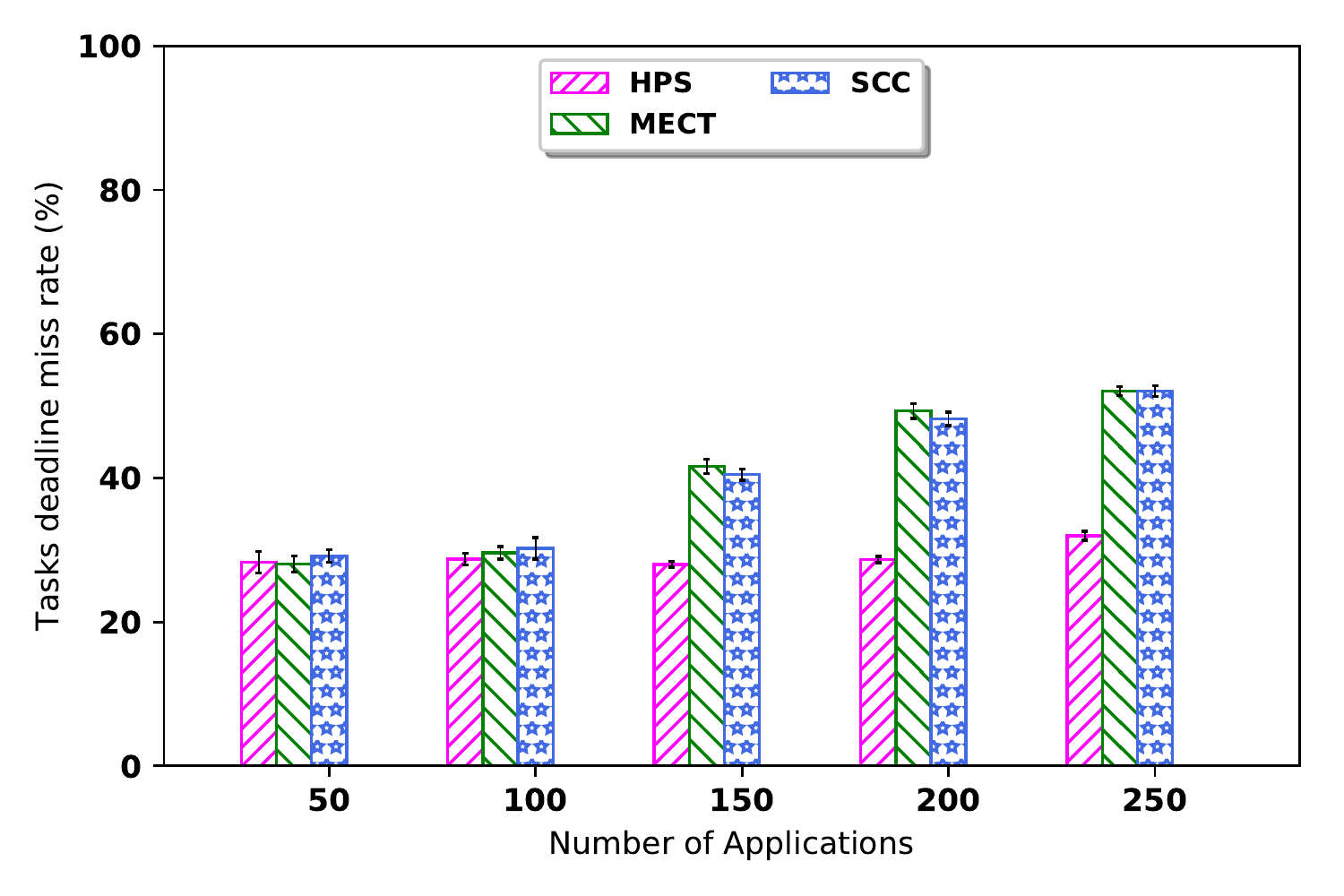}
	\vspace{-2mm}
	\caption{Increasing oversubscription level with increased number of applications.}
	\label{fig:oversubscriptionLevel}
\end{figure}
The result reflects that with the increased number of applications, the deadline miss rate gets increased for all of the heuristics. When the system starts getting oversubscribed for 150 applications the difference in performance of the heuristics is significant. Specifically, when the system is fully oversubscribed with 250 applications $HPS$ performs approximately 21\% better than $MECT$ and $SCC$. This is because our proposed heuristic considers both communication and computation overhead whereas other heuristics consider only one of them. 


\section{Related Work}\label{relatedwork}
The researcher has previously exploited the concept of edge computing for delay-tolerant networks. Lorenzo \etal in \cite{lorenzo2017robust} proposed an edge computing system with resource allocation design in the wireless network using mobile devices to mitigate the problem of network congestion. In\cite{chang2018adaptive} Chang \etal proposed an optimized resource migration scheme from mobile IoT devices to heterogeneous Cloud-Fog-Edge computing environment which considers hardware limitation of edge devices. The serverless computing concept in edge level is explored by Nastic and Dustdar in \cite{nastic2018towards}. Prior research works have been undertaken on resource allocation of edge computing systems with unreliable network connectivity~\cite{bp2}. However, these research works neither consider the heterogeneity of the edge resources nor have the self-organization and autonomy abilities~\cite{7926320,wang13,zhu17}. The specific problem of resource provisioning in serverless edge federation with low-connectivity to the back-end datacenters has not been explored in the context of remote smart oil fields. Efforts towards smart oil fields have been predominantly focused on analyzing the big data extracted from oil wells~\cite{cameron14}, applying machine learning methods to reduce exploration and drilling costs~\cite{mehdi2017}, or warning systems for early prediction of disasters~\cite{xu17}. These solutions are all reliant on onshore datacenters, which is not viable for remote and offshore oil fields~\cite{bogaert2004Offshore}.



\section{Conclusion}
In this paper, we propose federation of serverless edge computing systems to enable green oil extraction. The system utilizes an efficient resource allocation scheme to cope with the uncertainties that exist in remote offshore smart oil fields. We leveraged the edge federation to utilize underutilized neighbor edge devices and historical data to predict the completion time of an incoming service request (task) with a probabilistic model. The model is aware of resource constraint nature of edge devices as well as the uncertainties due to stochastic nature in communication. Both of the uncertainties were incorporated while predicting the probability of success within a strict deadline period of task completion. Experimental results demonstrate that our proposed model can improve the completion rate of urgent services compared to other conventional models. Simulation result reflects that for increasing service load, our proposed heuristic outperformed (up to 21\%) the baseline heuristics. The future plan of this work is to utilize approximate computing in serverless edge to improve the performance.

\bibliographystyle{ieeetr}
\balance
\bibliography{references}

\begin{thebibliography}{10}

\bibitem{mathieson2007forces}
D.~Mathieson {\em et~al.}, ``Forces that will shape intelligent-wells
  development,'' {\em Journal of Petroleum Technology}, vol.~59, pp.~14--16,
  Aug. 2007.

\bibitem{White20303}
H.~K. White, P.-Y. Hsing, W.~Cho, T.~M. Shank, E.~E. Cordes, A.~M. Quattrini,
  R.~K. Nelson, R.~Camilli, A.~W.~J. Demopoulos, C.~R. German, J.~M. Brooks,
  H.~H. Roberts, W.~Shedd, C.~M. Reddy, and C.~R. Fisher, ``Impact of the
  deepwater horizon oil spill on a deep-water coral community in the gulf of
  mexico,'' {\em Proceedings of the National Academy of Sciences}, vol.~109,
  pp.~20303--20308, Feb. 2012.

\bibitem{sof2}
\text{Cisco}, ``{A New Reality for Oil \& Gas: Data Management and
  Analytics},'' White paper, April 2015.

\bibitem{sof3}
WoodMackenzie, ``Why are some deepwater plays still attractive?,'' White paper,
  Sep. 2017.

\bibitem{sof5}
\text{Gartner}, ``Top 10 technology trends impacting the oil and gas industry
  in 2015,'' White paper, Mar. 2015.

\bibitem{sof8}
T.~Nuth, ``{Challenges of Networking the Smart Oil Field},'' in {\em Automation
  World}, Feb. 2016.

\bibitem{sof10W}
\text{Motorola}, ``{Improving Safety and Productivity in Oil and Gas
  Operations},'' White paper 2014.

\bibitem{sof11W}
\text{Motorola}, ``{Top Six Priorities for Team Communications},'' White paper,
  May 2016.

\bibitem{sof13W}
\text{ABB}, ``{Field Area Communication Networks for Digital Oil and Gas
  Fields},'' White paper, October 2014.

\bibitem{sof15W}
\text{Huawei}, ``{Exploring the Needs of the Digital Oilfield},'' White paper
  2012.

\bibitem{sof1}
S.~Prabhu, E.~Gajendran, and N.~Balakumar, ``Smart oil field management using
  wireless communication techniques,'' {\em International Journal of Inventions
  in Engineering \& Science Technology}, pp.~2454--9584, Jan. 2016.

\bibitem{LIU20112611}
P.~Liu, X.~Li, J.~J. Qu, W.~Wang, C.~Zhao, and W.~Pichel, ``Oil spill detection
  with fully polarimetric uavsar data,'' {\em Marine Pollution Bulletin},
  vol.~62, pp.~2611 -- 2618, Dec. 2011.

\bibitem{sof12W}
N.~G. Franconi, A.~P. Bunger, E.~Sejdić, and M.~H. Mickle, ``Wireless
  communication in oil and gas wells,'' {\em Energy Technology}, vol.~2,
  pp.~996--1005, Oct. 2014.

\bibitem{giacobbe2017internet}
M.~Giacobbe, R.~Pietro, A.~Zaia, and A.~Puliafito, ``The internet of things in
  oil and gas industry: A multi criteria decision making brokerage strategy,''
  in {\em Proceedings of the 4th International Conference on Automation,
  Control Engineering and Computer Science (ACECS)), Proceedings of Engineering
  and Technology-PET}, vol.~21, pp.~47--52, Mar. 2017.

\bibitem{pickering2015adopting}
J.~Pickering, S.~Sengupta, M.~Pfitzinger, {\em et~al.}, ``Adopting cloud
  technology to enhance the digital oilfield,'' in {\em Proceedings of
  International Petroleum Technology Conference}, Dec. 2015.

\bibitem{cho2015safety}
J.~Cho, G.~Lim, T.~Biobaku, S.~Kim, and H.~Parsaei, ``Safety and security
  management with unmanned aerial vehicle (uav) in oil and gas industry,'' {\em
  Journal of Procedia Manufacturing}, vol.~3, pp.~1343--1349, Jul. 2015.

\bibitem{Deepwater}
E.~Horsch, E.~English, and J.~Stein, ``Lessons from the deepwater horizon oil
  spill on the use of aerial recreation surveys,'' {\em Journal of Survey
  Statistics and Methodology}, vol.~5, pp.~310--325, Sep. 2017.

\bibitem{gibson2017evaluating}
D.~Gibson, D.~H. Catlin, K.~L. Hunt, J.~D. Fraser, S.~M. Karpanty, M.~J.
  Friedrich, M.~K. Bimbi, J.~B. Cohen, and S.~B. Maddock, ``Evaluating the
  impact of man-made disasters on imperiled species: Piping plovers and the
  deepwater horizon oil spill,'' {\em Journal of Biological Conservation},
  vol.~212, pp.~48--62, May. 2017.

\bibitem{bp1}
S.~K. Pandey, K.-H. Kim, and K.-T. Tang, ``A review of sensor-based methods for
  monitoring hydrogen sulfide,'' {\em Journal of TrAC Trends in Analytical
  Chemistry}, vol.~32, pp.~87 -- 99, Feb. 2012.

\bibitem{toxicgas}
\text{David Riddle}, ``Danger and detection of hydrogen sulphide gas in oil and
  gas exploration and production,'' White paper, April 2009.

\bibitem{ismail2015Docker}
B.~I. Ismail, E.~M. Goortani, M.~B. Ab~Karim, W.~M. Tat, S.~Setapa, J.~Y. Luke,
  and O.~H. Hoe, ``Evaluation of docker as edge computing platform,'' in {\em
  Proceedings of the IEEE Confernece on Open Systems (ICOS)}, pp.~130--135,
  Aug. 2015.

\bibitem{PERRONS2013732}
R.~K. Perrons and A.~Hems, ``Cloud computing in the upstream oil \& gas
  industry: A proposed way forward,'' {\em Journal of Energy Policy}, vol.~56,
  pp.~732 -- 737, Jan. 2013.

\bibitem{reza2010applications}
M.~reza Akhondi, A.~Talevski, S.~Carlsen, and S.~Petersen, ``Applications of
  wireless sensor networks in the oil, gas and resources industries,'' in {\em
  Proceedings of the 24th IEEE International Conference on Advanced Information
  Networking and Applications (AINA)}, pp.~941--948, Apr. 2010.

\bibitem{salehi2016stochastic}
M.~A. Salehi, J.~Smith, A.~A. Maciejewski, H.~J. Siegel, E.~K. Chong,
  J.~Apodaca, L.~D. Brice{\~n}o, T.~Renner, V.~Shestak, J.~Ladd, {\em et~al.},
  ``Stochastic-based robust dynamic resource allocation for independent tasks
  in a heterogeneous computing system,'' {\em Journal of Parallel and
  Distributed Computing (JPDC)}, vol.~97, pp.~96--111, Jun. 2016.

\bibitem{5GQualcom}
\text{Gabriel Brown}, ``{Exploring 5G New Radio: Use cases, Capabilities \&
  Timeline},'' White paper, Qualcomm, Sep. 2016.

\bibitem{avigad2017formally}
J.~Avigad, J.~H{\"o}lzl, and L.~Serafin, ``A formally verified proof of the
  central limit theorem,'' {\em Journal of Automated Reasoning}, vol.~59,
  pp.~389--423, Feb. 2017.

\bibitem{cheadle2003analysis}
C.~Cheadle, M.~P. Vawter, W.~J. Freed, and K.~G. Becker, ``Analysis of
  microarray data using z score transformation,'' {\em The Journal of molecular
  diagnostics}, vol.~5, pp.~73--81, Dec. 2003.

\bibitem{knezevic2008overlapping}
A.~Knezevic, ``Overlapping confidence intervals and statistical significance,''
  {\em StatNews: Cornell University Statistical Consulting Unit}, vol.~73, Oct.
  2008.

\bibitem{sonmez2017edgecloudsim}
C.~Sonmez, A.~Ozgovde, and C.~Ersoy, ``Edgecloudsim: An environment for
  performance evaluation of edge computing systems,'' in {\em Proceedings of
  the 2nd International Conference on Fog and Mobile Edge Computing (FMEC)},
  FMEC '17, pp.~39--44, May 2017.

\bibitem{calheiros2011cloudsim}
R.~N. Calheiros, R.~Ranjan, A.~Beloglazov, C.~A. De~Rose, and R.~Buyya,
  ``Cloudsim: a toolkit for modeling and simulation of cloud computing
  environments and evaluation of resource provisioning algorithms,'' {\em
  Software: Practice and experience}, vol.~41, pp.~23--50, Jan. 2011.

\bibitem{goyal1998analysis}
R.~Goyal, S.~Kota, R.~Jain, S.~Fahmy, B.~Vandalore, and J.~Kallaus, ``Analysis
  and simulation of delay and buffer requirements of satellite-atm networks for
  tcp/ip traffic,'' {\em arXiv preprint cs/9809052}, Mar. 1998.

\bibitem{khemka2015utility}
B.~Khemka, R.~Friese, S.~Pasricha, A.~A. Maciejewski, H.~J. Siegel, G.~A.
  Koenig, S.~Powers, M.~Hilton, R.~Rambharos, and S.~Poole, ``Utility
  maximizing dynamic resource management in an oversubscribed
  energy-constrained heterogeneous computing system,'' {\em Journal of
  Sustainable Computing: Informatics and Systems}, vol.~5, pp.~14--30, Aug.
  2015.

\bibitem{khemka2015functions}
B.~Khemka, R.~Friese, L.~D. Briceno, H.~J. Siegel, A.~A. Maciejewski, G.~A.
  Koenig, C.~Groer, G.~Okonski, M.~M. Hilton, R.~Rambharos, {\em et~al.},
  ``Utility functions and resource management in an oversubscribed
  heterogeneous computing environment,'' {\em Journal of IEEE Transactions on
  Computers}, vol.~64, pp.~2394--2407, Aug. 2015.

\bibitem{hussainrobust}
R.~F. Hussain, M.~A. Salehi, A.~Kovalenko, S.~Salehi, and O.~Semiari, ``Robust
  resource allocation using edge computing for smart oil fields,'' in {\em
  Proceedings of the 24th International Conference on Parallel and Distributed
  Processing Techniques \& Applications}, Aug. 2018.

\bibitem{lorenzo2017robust}
B.~Lorenzo, J.~Garcia-Rois, X.~Li, J.~Gonzalez-Castano, and Y.~Fang, ``A robust
  dynamic edge network architecture for the internet-of-things,'' {\em Journal
  of IEEE Network}, vol.~32, pp.~8--15, Jan. 2017.

\bibitem{chang2018adaptive}
C.~Chang, A.~Hadachi, and S.~Srirama, ``Adaptive edge process migration for iot
  in heterogeneous cloud-fog-edge computing environment,'' {\em arXiv preprint
  arXiv:1811.10939}, Nov. 2018.

\bibitem{wunderlich2017network}
S.~Wunderlich, J.~A. Cabrera, F.~H. Fitzek, and M.~Reisslein, ``Network coding
  in heterogeneous multicore iot nodes with dag scheduling of parallel matrix
  block operations,'' {\em Journal of IEEE Internet of Things}, vol.~4,
  pp.~917--933, May 2017.

\bibitem{wang2013nested}
Y.~Wang, X.~Lin, and M.~Pedram, ``A nested two stage game-based optimization
  framework in mobile cloud computing system,'' in {\em Proceedings of IEEE 7th
  International Symposium on Service Oriented System Engineering (SOSE)},
  pp.~494--502, Mar. 2013.

\bibitem{hu2016green}
Z.~Hu, Y.~Wei, X.~Wang, and M.~Song, ``Green relay station assisted cell
  zooming scheme for cellular networks,'' in {\em Proceedings of 12th
  International Conference on Natural Computation, Fuzzy Systems and Knowledge
  Discovery}, pp.~2030--2035, Aug. 2016.

\bibitem{xiaoyu2012smart}
X.~Xiaoyu, L.~Yan, F.~Zhong, J.~Wang, D.~Xu, F.~Wang, L.~Gan, J.~Zhou, Z.~Lin,
  W.~Pan, {\em et~al.}, ``Smart well technology in daqing oil field,'' in {\em
  Proceedings of the Abu Dhabi International Petroleum Conference and
  Exhibition}, Nov. 2012.

\bibitem{al2008intelligent}
R.~Al-Zahrani, I.~H. Al-Arnaout, S.~Jacob, Z.~A. Rahman, {\em et~al.},
  ``Intelligent wells to intelligent fields: Remotely operated smart well
  completions in haradh-iii,'' in {\em Proceedings of the Intelligent Energy
  Conference and Exhibition}, Feb. 2008.

\bibitem{jose2016smart}
S.~Jose, T.~Sookdeo, I.~Says, Y.~Carreno, R.~Ferrero, {\em et~al.}, ``The smart
  digital platform for effective well control management,'' in {\em Proceedings
  of the Society of Petroleum Engineers (SPE) Intelligent Energy International
  Conference and Exhibition}, Sep. 2016.

\bibitem{oliver2018enhancing}
P.~R. Oliver {\em et~al.}, ``Enhancing operations with digital technology to
  bring expertise to site,'' in {\em Proceedings of Offshore Technology
  Conference Asia}, Mar. 2018.

\bibitem{aggrey6data}
D.~D. R. A. A. A. K. M.~R. Aggrey, George~Hayford, ``Data richness and
  reliability in smart-field management - is there value?,'' in {\em
  Proceedings of SPE Annual Technical Conference and Exhibition}, Sep. 2006.

\bibitem{ekebafe12}
A.~Ekebafe, Abraham;~Ogan, ``Smart well technology application in deepwater
  field development,'' in {\em Proceedings of Nigeria Annual International
  Conference and Exhibition}, Aug. 2012.

\bibitem{van12smartfield}
F.~G. Van~den Berg, ``Smart fields - optimising existing fields,'' in {\em
  Proceedings of Digital Energy Conference and Exhibition}, Aug. 2007.

\bibitem{parapuram2017prediction}
G.~K. Parapuram, M.~Mokhtari, J.~B. Hmida, {\em et~al.}, ``Prediction and
  analysis of geomechanical properties of the upper bakken shale utilizing
  artificial intelligence and data mining,'' in {\em Proceedings of the
  SPE/AAPG/SEG Unconventional Resources Technology Conference, Austin, TX,
  USA}, pp.~24--26, Jul. 2017.

\bibitem{cameron2014big}
D.~Cameron {\em et~al.}, ``Big data in exploration and production: Silicon
  snake-oil, magic bullet, or useful tool?,'' in {\em Proceedings of SPE
  Intelligent Energy Conference \& Exhibition}, Apr. 2014.

\bibitem{skedsmo2016oil}
M.~Skedsmo, R.~Ayasse, N.~Soleng, M.~Indregard, {\em et~al.}, ``Oil spill
  detection and response using satellite imagery, insight to technology and
  regulatory context,'' in {\em Proceedings of the Society of Petroleum
  Engineers (SPE) International Conference and Exhibition on Health, Safety,
  Security, Environment, and Social Responsibility}, Apr. 2016.

\bibitem{prabhu2017smart}
B.~Prabhu, E.~Gajendran, and N.~Balakumar, ``Smart oil field management using
  wireless communication techniques,'' Jan. 2017.

\end{thebibliography}




\end{document}